\documentclass[runningheads]{svmult}

\usepackage{makeidx}   % allows index generation
\usepackage{graphicx}  % standard LaTeX graphics tool
\usepackage{subeqnar}  % subnumbers individual equations
\usepackage{multicol}  % used for the two-column index
\usepackage{physprbb}  % modified textarea for proceedings,
\makeindex             % used for the subject index
\begin{document}
\title*{ESO for GOODS' sake}

\titlerunning{ESO Plans for GOODS}
% allows abbreviation of title, if the full title is too long
% to fit in the running head
%
\author{A. Renzini$^1$, C. Cesarsky$^1$, S. Cristiani$^2$, L. da Costa$^1$, 
R. Fosbury$^2$, R. Hook$^2$, B. Leibundgut$^1$, P. Rosati$^1$, \& B. 
Vandame$^1$
}
\authorrunning{Alvio Renzini and the ESO/GOODS Team} 
% if there are more than two authors,
% please abbreviate author list for running head
%
%
\institute{$^1$ESO, 85748 Garching b. M\"unchen, Germany\\
           $^2$ST-ECF, 85748 Garching b. M\"unchen, Germany}
\maketitle              % typesets the title of the contribution

\begin{abstract}
Currently public ESO data sets pertinent to the CDFS/GOODS field are briefly
illustrated along with an indication on how to get access to them. Future ESO 
plans for complementing the GOODS database with optical/IR imaging and optical
spectroscopy are also described.

\end{abstract}

\section{Introduction}
Recognizing the unique and long-lasting scientific value of the
``SIRTF Legacy Programs'', the Director General of ESO issued an
``Open Letter to the ESO Community'' to ensure ``adequate coverage
from the observatories at La Silla and Paranal for all the approved
Legacy Programs which have a substantial participation from the ESO
community. ESO will ensure that appropriate allocation of time on
relevant instruments, in line with the scientific goals of approved
SIRTF Legacy programmes, is made in a timely manner. In the spirit of
all the Legacy Programs, the resulting data will be immediately made
public worldwide.'' (see
http://www.eso.org/observing/misc/20000824.sirtf.html).  In this
spirit, a team of ESO astronomers joined forces with the North
American team led by Mark Dickinson at STScI, and the GOODS SIRTF
Legacy Proposal was submitted in 2000, then complemented in 2001 by
the HST/ACS Treasury Proposal led by Mauro Giavalisco. 
Both projects are now being implemented. The
accompanying paper by Dickinson \& Giavalisco describes the main
scientific goals of the GOODS project, along with the planned
observations with SIRTF and HST. 
This paper complements it
with a description of the data being provided by ESO, and of the
possibilities of spectroscopic follow up with the VLT. Involvement of the 
ESO community in the planning of the observations as well as in the reduction 
effort will be actively pursued.

\section{Optical and Near-IR Imaging}

There are several imaging data sets from observations taken at ESO
telescopes that include the CDFS/GOODS field and are already publicly
available. Some are part of the public ESO Imaging Survey (EIS), and some
were obtained as part of ``private'' projects, but the data have
become available after the one year proprietary period has expired.
Other data have been obtained specifically for the GOODS project as a 
public survey (VLT
Large Programme 68.A-0485, PI C. Cesarsky). Table 1 summarizes the
situation as of February 2002. Fig. 1 shows the CDFS/GOODS
field with superimposed the {\it brickwall} mosaic of the VLT/ISAAC
observations currently under way. The raw and calibration $JHK$ data
having the planned integration time are already available for those bricks
which identification number has the large format in Fig. 1. $JK$ data for
bricks 10, 11, 15 
and 16 were secured by
ESO programmes 64.O-0643, 66.A-0572 and 68.A-0544, whose PI
(E. Giallongo) has kindly agreed to waive the residual proprietary
time in such a way to have the reduced data being publicly released at
the same time as the reduced GOODS data. For these bricks GOODS also provides 
the $H$-band data. The reduction of 
coadded, flux and astrometrically calibrated data is underway by the
EIS Team, with the data release being planned for the end of March,
2002. The first public release of mosaiced images and source catalogs
is planned for the end of July, 2002. ISAAC observations to complete
the whole brickwall will continue in 2002 and 2003 as part of the
GOODS VLT Large Programme. Real time information on the progress of
the ESO/GOODS observations can be obtained from the URL
http://www.eso.org/science/goods/, with links to the EIS and other databases.

While deep optical imaging was initially also envisaged with the VLT, this has 
become partly redundant after the approval of the GOODS/ACS Treasury Program,
which will provide deep $BViz$ imaging.
Exception refers to deep $U$-band imaging with
the VIMOS instrument and/or with a UV-optimized FORS-1 (if and when it 
becomes available). Depending on the public availability of data, additional 
VIMOS imaging may or may not be also needed
for the selections of spectroscopic targets in the field surrounding the
GOODS field (see Section 3.1).
\begin{figure}[h]
\begin{center}
\includegraphics[width=1.0\textwidth]{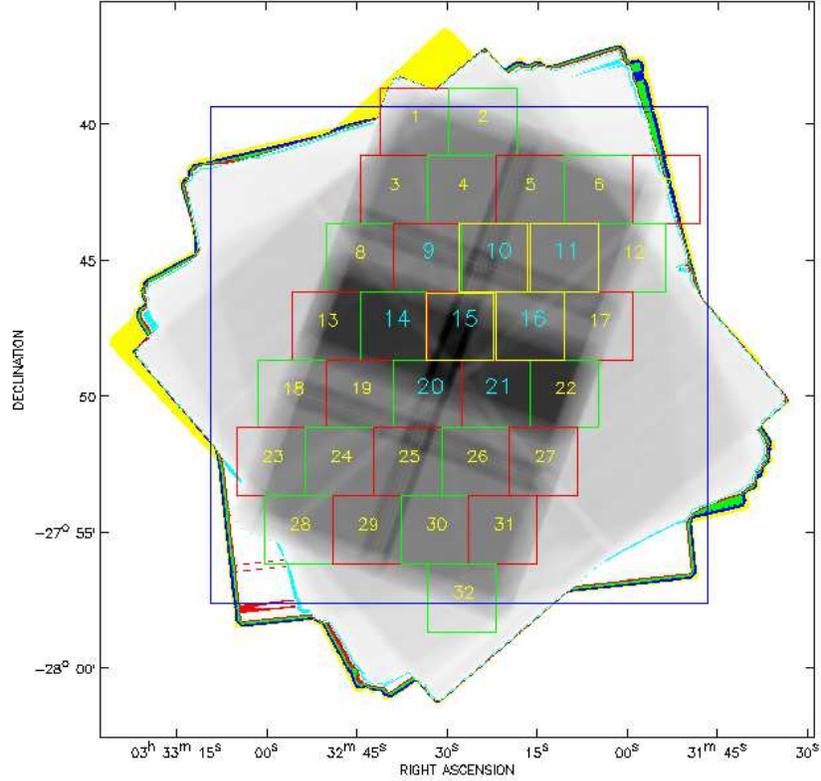}
\end{center}
\caption[]{The ISAAC mosaic coverage of the CDFS/GOODS field (the shaded IRAC 
exposure map) is shown by the
{\it brickwall} with individual pointings numbered from 1 to 32. Bricks with 
a large-size identification number have already been completed. The Chandra 
coverage is shown  by the 
set of shaded squares with various orientations. The centered, unshaded square
shows the EIS/DPS Deep-2c field as covered by SOFI.}
\label{eps3}
\end{figure}

\begin{table}
\caption{ESO imaging data set on CDFS/GOODS (February 2002)}
\begin{center}
\renewcommand{\arraystretch}{1.4}
\setlength\tabcolsep{5pt}
\begin{tabular}{llllll}
\hline\noalign{\smallskip}
Bands     &   Tel./Instrument  & Area &  5-$\sigma$ AB mag limits & Programme\\
\noalign{\smallskip}
\hline
\noalign{\smallskip}
$UBVRI$ & 2.2m/WFI & $30'\times 30'$ &$U<26.0,\; B<26.4,\; V<25.4$; &EIS/DPS\\
        &          &                 &$R<25.5;\; I<24.7$            &       \\ 
\hline
$JK_{\rm s}$ &NTT/SOFI & $20'\times 20'$ &$J<23.4;\; K_{\rm s}<22.6$ &EIS/DPS\\
\hline
\hline
$RI$         &VLT/FORS & $15'\times 15'$ & $R<27.0;\; I<26.0$ & 64.O-0621 \\
\hline
\hline
$JHK_{\rm s}$ &VLT/ISAAC$^*$ &$10'\times 16'$  &$J<25.3;\; H<24.8;\; K_{\rm 
    s}<24.4$ & GOODS \\
\hline
$BVR$ & 2.2m/WFI & $30'\times 30'$ &$B<27.0;\; V< 26.5;\; R<26.5$ & GOODS \\
\hline
\hline
$^*$ See text.
\end{tabular}
\end{center}
\label{Tab1}
\vskip -1.0 truecm
\end{table}

\begin{figure}[h]
\begin{center}
\hbox{
{\includegraphics[width=0.5224\textwidth]{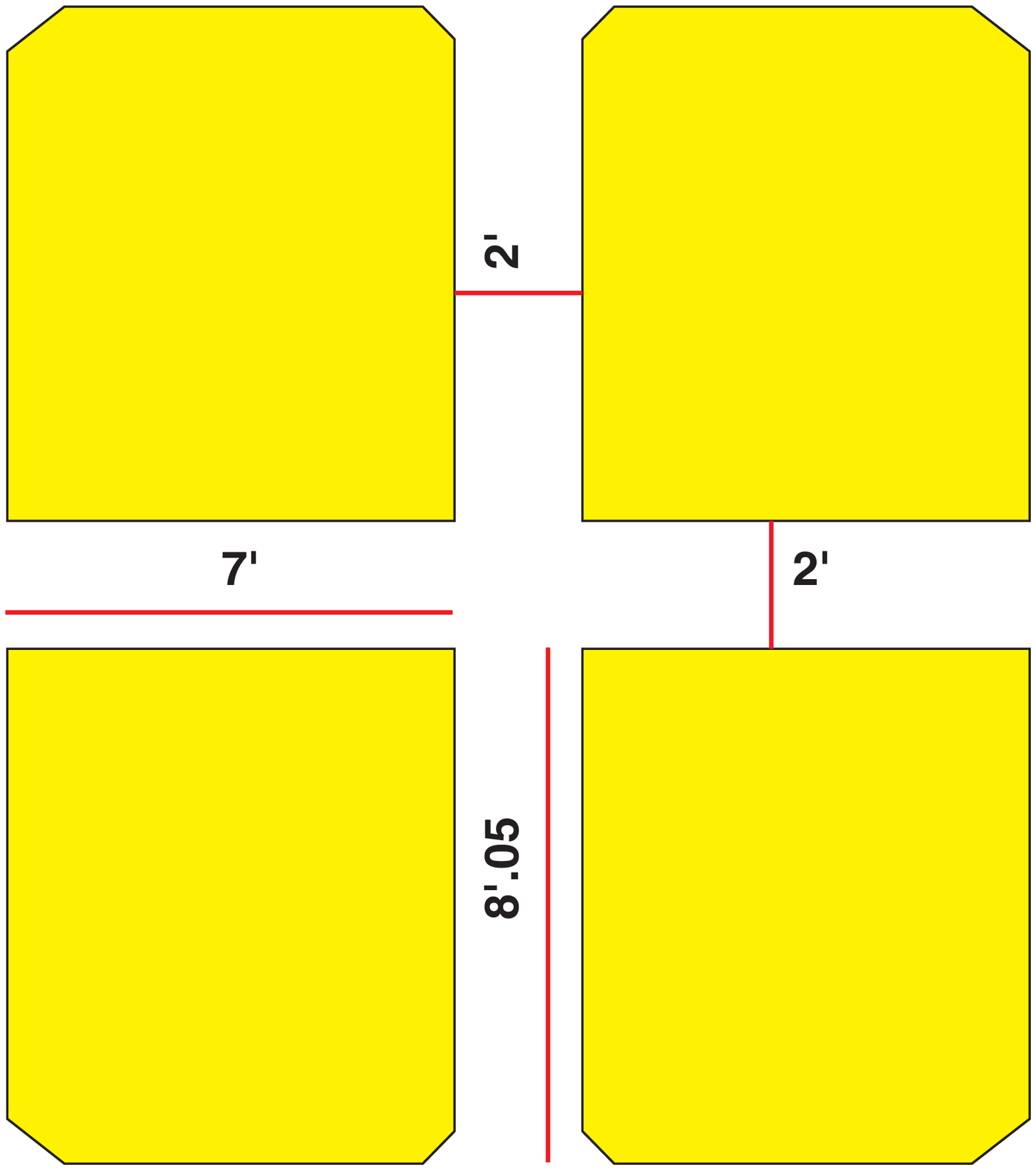}}
\hspace{0.0850\textwidth}
{\includegraphics[width=0.3628\textwidth]{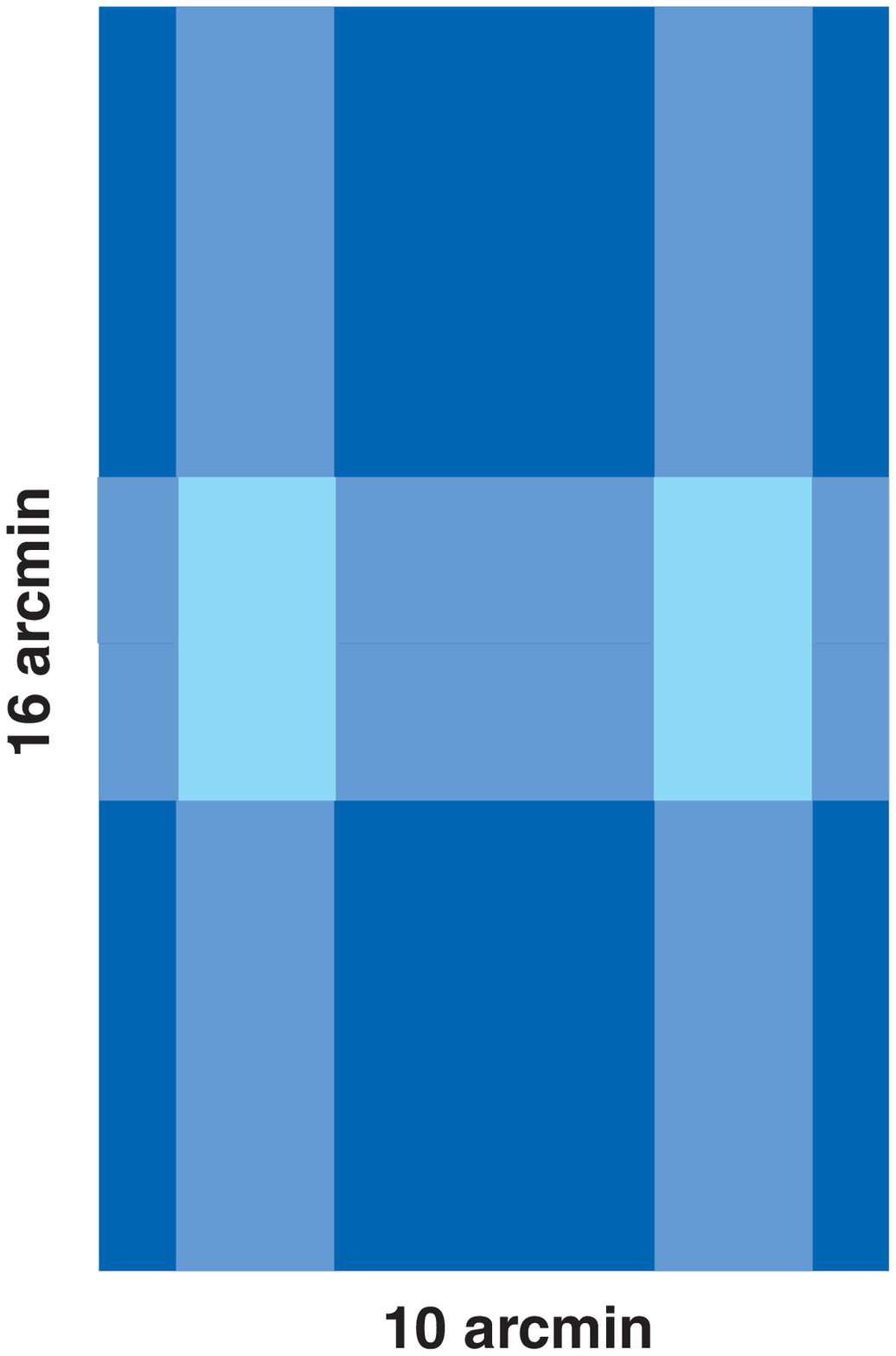}}
}
\end{center}
\caption[]{{\sl Left}: the geometry of the VIMOS field of view, for either 
imaging or multiobject spectroscopy. \quad
{\sl Right}: the CDFS/GOODS $10'\times 16'$ field, in the same scale as the
VIMOS FoV (left). The VIMOS imprinting on the GOODS field is rendered 
with different shadings. Dark areas: areas common to
all 4 default VIMOS pointings (A set of subfields). Shaded areas: areas 
common to 2 default VIMOS pointings (B set of subfields. Lightly shaded areas:
areas covered by only one VIMOS pointing (C set of subfields).
}
\label{eps2}
\end{figure}
\section{VLT Spectroscopy}

The most important ESO contribution to the GOODS project will
certainly be the wide spectroscopic coverage of the field. 
No other Southern observatory has high-multiplex spectroscopic capabilities
comparable to those of the VLT.
As of
February 2002, more than 30 nights of multi-object spectroscopy have already
been allocated at VLT/FORS1-2 for several scientific programs (e.g. the 
$K<20$ redshift survey; identification of Chandra sources; study of high-z
dropouts). To date, several hundred redshifts have being measured in
the CDF-S area.  This increasing spectroscopic data set amassed with the
VLT is already public in the ESO Science Archive or will become public
before the SIRTF data are taken.

\begin{table}
\caption{Cumulative Source Counts in the CDFS Field}
\begin{center}
\renewcommand{\arraystretch}{1.2}
\setlength\tabcolsep{5pt}
\begin{tabular}{lllll}
\hline\noalign{\smallskip}
$M_{\rm AB}$ & $N^{\rm EIS}_{\rm R}$ & $N^{\rm EIS}_{\rm I}$ & $N^{\rm 
          GOODS}_{\rm R}$ & $N^{\rm GOODS}_{\rm I}$ \\
\noalign{\smallskip}
\hline
\noalign{\smallskip}
22 &     3628 &   5794  &     600  &   1000 \\
23 &     8009 &  11096  &    1300  &   2000 \\
24  &   17962 &  20755  &    3000  &   3500 \\
25  &   35949 &  31148* &    6000  &   5500$^*$ \\
26  &   52840$^*$&      &    9400$^*$  &        \\
\hline
\end{tabular}
\end{center}
\label{Tab2}
\hskip 3 truecm $^*$ Incomplete counts in these magnitude bins.
\end{table}

However, these data either cover only a small fraction of the GOODS
field, or only very specific types of targets, such as
e.g. X-ray sources or a few Lyman-break galaxies. Table 2 gives the
cumulative source number counts in the $R$ and $I$ bands (columns 2
and 3, respectively) as derived from the EIS/DPS data for the Deep-2c
field that includes the CDFS/GOODS field (Arnouts et al. 2002, A\&A, 379,
740).
Scaling by the areas, columns 4 and 5 give the number of sources
within the GOODS field.

It appears from Table 2 that no more than $\sim 6000$ objects in
the GOODS field are bright enough (i.e. down to $R_{\rm AB}$ or 
$I_{\rm AB}=25$)  for a
low-resolution spectrum to be useful, e.g. to provide the
redshift. Hence, the multiplex capabilities of the VLT instruments ensure
the possibility to observe them all in a quite reasonable
amount of telescope time. This is now explored in more quantitative
terms.
\subsection{VIMOS Spectroscopy}
The layout of the VIMOS FoV is shown in Fig. 2 (left panel).  
A minimum of 4 VIMOS
pointings is necessary to cover the whole GOODS field.  A default
pattern may consist of 4 VIMOS pointings in which in turn each of the 4
outer corners of the VIMOS FoV coincides with each of the 4 GOODS
field corners, with the long (dispersion)
axis of VIMOS parallel to the long side of GOODS.  The missing
triangles at the VIMOS corners are ignored.

For each pointing, the central cross gaps of VIMOS separate 4
sub-fields over GOODS, for a total coverage of 112 arcmin$^2$. Hence,
for each pointing, a fraction 112/160 (70\%) of the GOODS field is
accessible to spectroscopy. The fraction of the VIMOS multiplex
expendable on GOODS is 112/224 (50\%,), i.e., with an {\it effective
multiplex} $0.5\times 800=400$ (in low resolution mode).

Moreover, the 4 default pointings of VIMOS imprint on the GOODS field a set of
subfields, with some being common to all 4 pointings, some only to 2
pointings, and 2 subfields are covered by only one pointing. One
defines these 3 sets of subfields as set A, B and C, respectively. The
area of set A and B is 72 arcmin$^2$ each (45\% of GOODS each) while
set C has an area of 16 arcmin$^2$ (10\% of GOODS).
The default (A,B,C) pattern imprinted on the GOODS field is shown in Fig. 2
(right panel).

The surface density of targets (6000/160=37.5 arcmin$^{-2}$) is 
$\sim 10$ times larger than the surface density of VIMOS slits (800/224=
3.6 arcmin$^{-2}$). Hence, if one insists on having completeness in set C
one needs at least 10 pointings to cover this small ($10\%$) part of the GOODS
field, while automatically set B and A would be observed two times and four
times more than strictly needed to ensure completeness, respectively. 

In a more time-saving approach, one may ensure competeness in set B,
which requires this set to be covered by at least 12 pointings.
Automatically, set C will be covered by 6 pointings (60\%
completeness), and set A by 24 pointings (twice more than strictly needed
to ensure completeness).  Hence, after having achieved completeness in
set A, the 12 additional pointings will allow the observation of a
subset of the targets in this field for longer integration times. In
the limit, some targets could be observed for up to $\sim 13$ times
the basic exposure time. In this scheme, 24 pointings are required and
assuming 4 hours integration time per pointing this makes a total of
$24\times 4$=96 hours of integration, plus 20\% overhead makes $\sim
120$ hours, or $\sim 13$ nights. These 13 nights will ensure that
$\sim 96\%$ of the targets are observed at least once with 4 hours
integration, while 45\% of the targets (those in set A) could be
observed with an integration of 8 hours each, or a smaller fraction
with even longer integrations.

This somewhat laborious exercise demonstrates that the geometries of the 
GOODS field 
and the VIMOS FoV combine in such a way that, when ensuring the complete 
spectroscopic coverage of the field, 
one has automatically the opportunity to integrate on some of the targets for
longer times than others. For this to be achieved some 
targets will have to be kept in more than one VIMOS multislit mask.

Very simple color criteria can be adopted to assign targets to either
observations with the red or the blue low-resolution grism of VIMOS,
in such a way to maximize the chance to obtain a high S/N spectrum. Similarly,
just a magnitude limit criterion can be adopted for the selection of targets 
to be observed for multiples of the basic exposure time.
Finally, having only 50\% of
the VIMOS spectroscopic multiplex used on the GOODS field, 
the other 50\% will be available to get spectra of objects in the accessible
area around GOODS. This should allow to get spectra of nearly as many objects
outside GOODS as inside it, i.e. ~6000 objects.
\subsection{FORS2 Spectroscopy}

While having a smaller FoV and lower multiplex compared to VIMOS,
FORS2 offers an attractive complementary capability.  Its throughput 
is about twice that of VIMOS longward of $\sim 800$ nm, with
the additional advantage of providing virtually fringe-free CCD
frames. On red objects, FORS2 may go $\sim 1$ mag or more deeper than
VIMOS, or could reach much better S/N, given also the higher
resolution ($\times 3$) that should help resolve the OH lines.  For
multiobject spectroscopy, the effective FoV of FORS2 is $\sim
6.8\times 3.2\simeq 22$ arcmin$^2$ and 9 pointing are sufficient to
cover the whole GOODS field, with minimal overlap between
pointings. With a multiplex $\sim 40$, such 9 pointings enables $\sim
360$ objects to be observed, selected according to a simple color 
criterion. 18 masks exposed for 4 hours each would permit observations 
of $> 500$ faint, red objects, with galaxies in faintest magnitude 
bin observed on two masks for a total of 8 hours.  The integration 
time for the faintest objects would be longer than for typical VIMOS 
exposures, ensuring that features will be recognizable in spectra 
with no emission lines.  This FORS2 program would require 
$9\times 2 \times 4\times 1.2=86$ hours of telescope time 
including overhead, or $\sim 10$ nights.

\section{Conclusions}

ESO is committed to provide its best possible contribution to the
scientific effort started with the GOODS SIRTF Legacy and HST/ACS
Treasury projects.  This will include complementary optical and
near-IR imaging, while a proposal will be submitted to the ESO
Observational Programme Committee aimed to provide the whole
astronomical community with the complete spectroscopic coverage of the
CDFS/GOODS field. This proposal will follow the lines sketched in this
paper, and should result in a complete, homogeneous database obtained with the
minimum possible VLT time. Together with the complementary data from space,
the ESO contribution is meant to establish a unique, 
long-lasting set of tools
for the study of galaxy formation and evolution, hence preparing the way
for further advances in the next decade with ALMA and NGST.

\end{document}